\documentstyle[graphics]{mn}
\newif\ifAMStwofonts
\input psfig.sty

\def\ao{A0538--66}
\def\exo{EXO\thinspace0531--66}
\def\ho{H\thinspace0544--665}
\def\til{$\sim$}
\def\msun{${\rm M}_{\odot}$}

\def\deg{$^{\circ}$}

\def\leq{\hbox{${_<\atop{\sim}}$}}

\def\ang{\thinspace\hbox{\AA}}
\def\ergsec{\thinspace\hbox{$\hbox{erg}\thinspace\hbox{s}^{-1}$}}

\def\arcsecdot{\nobreak\ifmmode{''\hskip-0.45em.\hskip0.08em}%
                         \else{$''\hskip-0.45em.\hskip0.08em$}\fi}
\def\arcsec{\nobreak\ifmmode{''\hskip-0.45em}%
                      \else{$''\hskip-0.45em$}\fi}

\title[MACHO photometry of EXO\thinspace0531--66 and H\thinspace0544--665]
      {MACHO Photometry of Two LMC Be X-ray Transients,
      EXO\thinspace0531--66 and H\thinspace0544--665}
\author[K.E. McGowan \& P.A. Charles]
{K.E. McGowan$^{1,2}$\thanks{email: mcgowan@lanl.gov} and P.A. 
Charles$^{2,3}$\\
$^{1}$Los Alamos National Laboratory, Los Alamos, NM 87545, USA \\
$^{2}$Department of Physics, University of Oxford, Oxford, OX1 3RH \\
$^{3}$Department of Physics \& Astronomy, University of Southampton, 
Southampton, SO17 1BJ}

\date{Accepted 
      Received       }

\pagerange{\pageref{firstpage}--\pageref{lastpage}}
\pubyear{2002}

\begin{document}

\maketitle

\label{firstpage}

\begin{abstract}
{Long-term variations are well-known in Be X-ray binaries, and are
attributed to non-orbital changes in the structure of the Be
circumstellar (equatorial) disc.  However, the timescales involved are
so long (tens of days, to years) that systematic studies have been
very restricted.  The \til8 year MACHO monitoring of the Large
Magellanic Cloud (LMC) therefore presents an ideal opportunity to
undertake such studies of Be X-ray systems that lie within the
monitored fields.  Here we present MACHO observations of two LMC Be
X-ray transients, \exo\ and \ho, the light curves of which show
substantial (\til0.5 mag) long-term variations.  However, our analysis
shows little evidence for any periodic phenomena in the light curves
of either source.  We find an upper limit for detection of a short
(1--100~d) periodicity in the {\it V}- and {\it R}-band light curves
of \exo\ of  0.041~mag and 0.047~mag semi-amplitude, respectively.
The upper limits for the {\it V}- and {\it R}-band data of \ho\ are
0.054~mag and 0.075~mag semi-amplitude, respectively.  Both \exo\ and
\ho\ become redder as they brighten, possibly due to variations in the
structure of the equatorial disc around the Be star.  Spectra of both
sources show H$\alpha$ emission; for \exo\ we find the emission varies
over time, thereby confirming its optical identification.}

\end{abstract}


\begin{figure*}
\begin{center}
%\leavevmode
\resizebox*{0.9\textwidth}{.37\textheight}{\rotatebox{-90}{\includegraphics{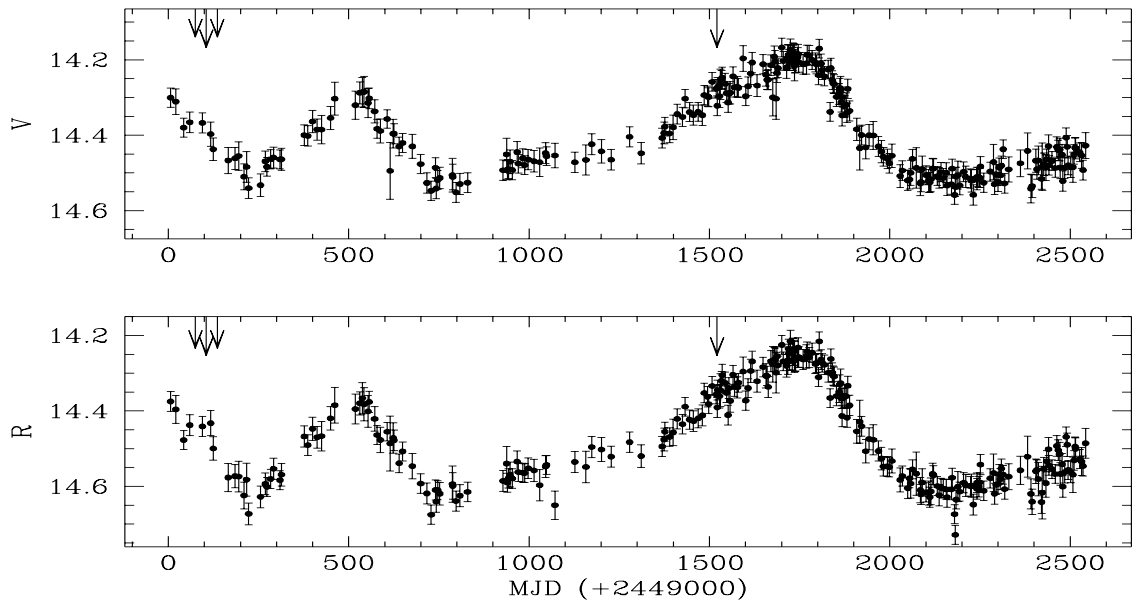}}}
\caption{The MACHO light curves in blue (top panel) and red (bottom
panel) filters of the proposed optical counterpart of \exo, which is
the northern component of a close double.  The {\it V} and {\it R}
magnitudes have been calculated using the absolute calibrations of the
MACHO fields.  The three arrows near day 100 in the top and bottom panels
indicate the observations taken during the X-ray outburst in 1993
(Haberl et al.\ 1995), the time of peak X-ray luminosity is indicated
by the middle arrow in each case.  The arrow near day 1500 in the top
and bottom panels indicates the time of the X-ray outburst observed by
Burderi et al.\ (1998).}\label{fig:exo_lc}
\end{center} 
\end{figure*}

\section{Introduction}

High mass X-ray binaries (HMXBs) generally fall into two subgroups,
supergiant HMXBs and Be X-ray binaries (Be/XRBs).  The companion star
in a Be/XRB is a Be (or Oe) star, with a typical mass of \til10--20
\msun.  A Be star is defined as a non-supergiant early-type star which
has at some time shown H$\alpha$ emission.   This Balmer emission,
along with a significant infrared excess, is believed to originate from
circumstellar material which forms an equatorial disc around the Be
star.  The Be star significantly underfills its Roche lobe, but is
thought to be rapidly rotating to explain the equatorial disc
(Slettebak 1987).  The orbital periods are long, ranging from \til15
days to years.

The light curves of Be/XRBs, and isolated Be stars, display
non-orbital long-term modulations which vary on timescales of tens of
days to years.  The modulations are thought to be due to a phase of
matter ejection from the star, the Be phenomenon (Slettebak 1987).
Due to the timescales involved it is difficult to study these
long-term modulations using normal observing programs.  Hence, the
observations obtained by the MACHO project provide a database from
which to investigate such systems.

In a previous paper (Alcock et al.\ 2001) we studied \til5 years of
MACHO observations of the Be X-ray transient (Be/XRT) \ao.  The
analysis revealed a long-term modulation of 421~d, together with the
previously known 16.6~d orbital period.  We attributed this long-term
period to the formation and depletion of the equatorial disc
surrounding the Be star.

These results motivated us to search for long and short term
periodicities in other Be/XRTs.  In this paper we present analysis of
the MACHO light curves of two other Be/XRTs, \exo\ and \ho.

\section{MACHO Observations}
\label{sect:macho}

The MACHO project monitored the LMC from 1992 July to 1999 December,
with the primary aim of detecting gravitational microlensing of
constituent LMC stars by intervening dark matter.  The observations
were made using the 1.27~m telescope at Mount Stromlo Observatory,
Australia.  A dichroic beamsplitter and filters provide simultaneous
CCD photometry in two passbands, a 'red' band (\til6300--7600 \ang)
and a 'blue' band (\til4500--6300 \ang).  The latter filter is a
broader version of the Johnson {\it V} passband (see Alcock et al.\
1995a, 1999 for further details).  The 'blue' and 'red' magnitudes
were transformed to Johnson {\it V} and Kron-Cousins {\it R}
respectively, using the absolute calibrations of the MACHO fields.

The images were reduced with the standard MACHO photometry code
{\scshape sodophot}, based on point-spread function fitting and
differential photometry relative to bright neighbouring stars.
Further details of the instrumental set-up and data processing may be
found in Alcock et al.\ (1995b, 1999), Marshall et al.\ (1994) and
Stubbs et al.\ (1993), and we note that the MACHO database is now in
the public domain (Allsman \& Axelrod 2001).

\begin{figure*}
\begin{center}
%\leavevmode
\resizebox*{.7\textwidth}{.32\textheight}{\rotatebox{-90}{\includegraphics{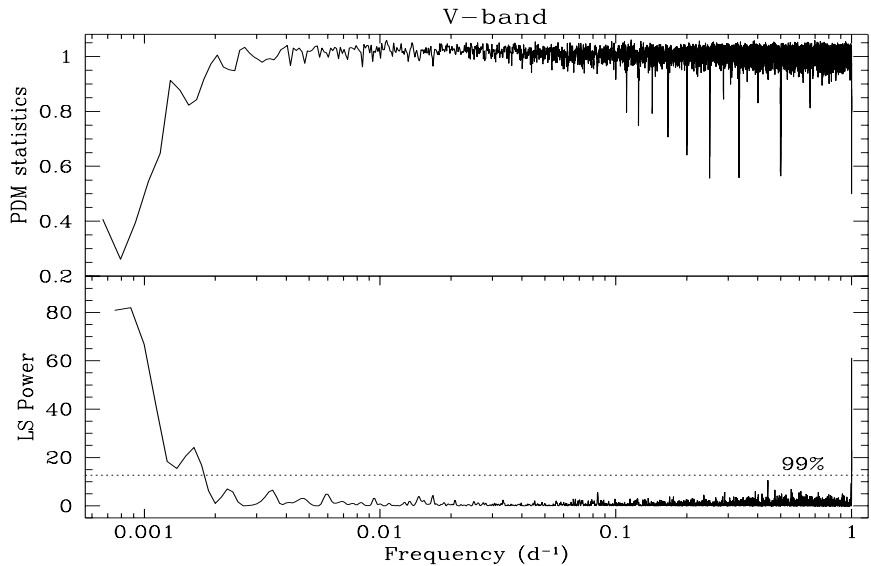}}}
\caption{PDM (top) and LS periodograms (bottom) for the {\it V}-band
data of \exo.  The frequency range and resolution is
6.67x$10^{-4}$--1.0 cycle d$^{-1}$ and 1.25x$10^{-4}$ cycle d$^{-1}$,
respectively.  The dotted line is the 99\% confidence
level.}\label{fig:exo_pdm}
\end{center} 
\end{figure*}

\section{EXO\thinspace0531--66}
\label{sect:exo}

\exo\ was discovered in 1983 during observations of the LMC~X--4
region with  {\it EXOSAT} (Pakull et al.\ 1985).  It was only seen for
a month at the end of 1983 (Pietsch, Rosso \& Dennerl 1989), and had
not been seen  in the {\it EINSTEIN} survey of the LMC (Long, Helfand
\& Grabelsky 1981).  The source was detected again in 1985 with the
SL2-XRT experiment on the Spacelab2 mission (Hanson et al.\ 1989).
The lack of detection of the source in {\it EXOSAT} observations made
between 1983 and 1985 indicates that \exo\ is a recurrent transient.

The proposed optical counterpart (Haberl, Dennerl \& Pietsch 1995) is
a Be star, which is the northern component of a close double.  During
a 4 year observing program of \exo\ with {\it ROSAT}, a strong
outburst was  detected in 1993 which lasted more than two months
(Haberl et al.\ 1995).  The intensity of this outburst was comparable
to that seen by {\it EXOSAT} in 1983, and a 13.7~s pulsation period
was detected (Dennerl, Haberl \& Pietsch 1996; Burderi et al.\ 1998),
indicating that \exo\ is a Be-neutron star binary.

Haberl et al.\ (1995) suggested that the observed outbursts were
consistent with being caused by the periastron passage of the neutron
star in a Be/XRB with an orbital period of 600--700~d.  However, they
could not rule out that the outbursts occur irregularly during a phase
of increased matter ejection from the Be star.  Dennerl et al.\ (1996)
constrained the orbital period to lie in the range of 4--70~d,
assuming that the rate of period change of the pulse period was
predominantly caused by Doppler shifts.  They found an orbital
solution with $P_{orb}$ = 25.4~d, $e$\til0.1 and $i$\til50\deg, for an
assumed mass for the compact object of 1.4 \msun, and 15 \msun\ for
the Be star companion.

The identification of the optical counterpart remains uncertain as
both the northern and southern stars in the close (\til4\arcsec\ )
double are early type stars.  Based solely on the X-ray error circle
(9\arcsec\ ),  either star could be the optical counterpart.  Stevens,
Coe \& Buckley (1999) obtained a spectrum of the northern component
which confirmed the presence of H$\alpha$ emission.  A low resolution
spectrum taken two nights later shows H$\beta$ also in emission.  No
spectrum has been taken of the southern component.

We searched the MACHO database for sources within a 9\arcsec\space\
radius of the X-ray position.  We find that the only source which
displays long-term variations typical of a Be star is the previously
proposed optical counterpart, the northern component of the double.

\subsection{Light curve}
\label{sect:exo_lc}

We show in Fig.~\ref{fig:exo_lc} the 'blue' and 'red' photometry of
the northern star proposed as the optical counterpart of \exo, from
1993 January 19 to 1999 December 30.  During three consecutive
observations of \exo, between March and May 1993, Haberl et al.\
(1995) found the source to be X-ray active.  The maximum observed
count rate occurred on April 27 1993 and was found to correspond to
2.0x$10^{36}$\ergsec\ in the {\it ROSAT} band (0.1--2.4~keV) for $d$ =
50 kpc.  The times of the three observations taken during the extended
outburst are marked on Fig.~\ref{fig:exo_lc}.  Unfortunately, the
MACHO project did not obtain observations of the source on these exact
dates.  However, observations were made near to these times and the
light curve shows a slight brightening of the source which could
indicate an optical response to the X-ray outburst.

Burderi et al.\ (1998) observed \exo\ on March 13-15 1997 with
BeppoSAX (also marked on Fig.~\ref{fig:exo_lc}).  The estimated X-ray
luminosity at the time of observation, converted into the {\it ROSAT}
band  (0.1--2.4~keV), was \til2.3x$10^{36}$\ergsec.  This value is
close to the luminosity of the outburst measured by Haberl et al.\
(1995).  In a low state the source is found to have a luminosity of
\leq $10^{35}$\ergsec\ (Haberl et al.\ 1995).  Burderi et al.\ (1998)
concluded that the source was in outburst at the time of their
observation.  During the \til2 day observation the light curve of
\exo\ in the medium energy band (1.0--10.5 keV) was relatively flat,
but with random fluctuations up to a factor of \til2 on timescales of
30~s.  The low energy light curve (0.1--1.8 keV) displayed flaring
episodes which were also observed in the high energy light curve
(15--60~keV).  The {\it V}- and {\it R}-band light curves show a
slight  brightening of the source at the time of the outburst reported
by Burderi et al.\ (1998), this could be related to the X-ray activity.

\subsection{Period Analysis}
\label{sect:exo_search}

Haberl et al.\ (1995) suggested that the orbital period for \exo\ was
in the range of 600--700~d.  However, Dennerl et al.\ (1996) proposed
a much shorter period in the range of 4--70~d, with a preferred value
of 25.4~d.  To investigate variability on these timescales we
performed a period search of the {\it V}- and {\it R}-band datasets
over frequency range 6.67x$10^{-4}$--1.0 cycle d$^{-1}$, with a
resolution of 1.25x$10^{-4}$ cycle d$^{-1}$.  Before the temporal
analysis was performed we detrended the data by subtracting a linear
fit.  We searched for periodicities in the light curves using a
Lomb-Scargle (LS) periodogram (Lomb 1976; Scargle 1982) and a phase
dispersion minimisation (PDM) periodogram (see Stellingwerf 1978).

\begin{figure*}
\begin{center}
%\leavevmode
\resizebox*{0.9\textwidth}{.37\textheight}{\rotatebox{-90}{\includegraphics{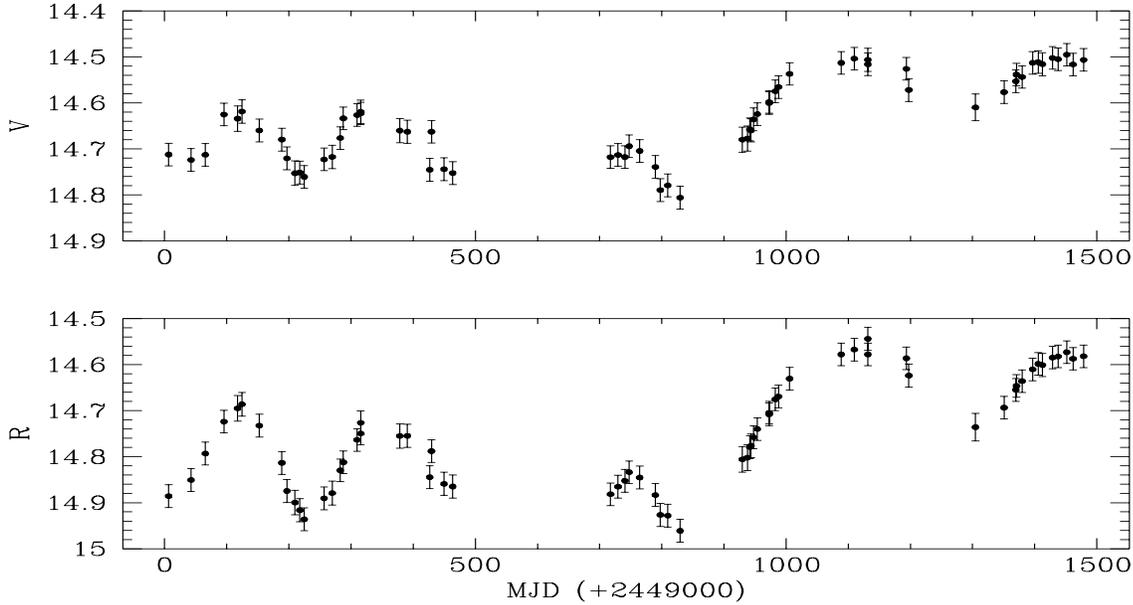}}}
\caption{The optical light curves in blue (top panel) and red (bottom
panel) filters of \ho\ from MACHO project observations.  The {\it V}
and {\it R}  magnitudes have been calculated using the absolute
calibrations of the MACHO  fields.}\label{fig:ho_lc}
\end{center} 
\end{figure*}

We show in Fig.~\ref{fig:exo_pdm} the period searches of the {\it
V}-band light curve of \exo, the results for the {\it R}-band data
were similar.  The largest peaks in the LS and PDM periodograms occur
at \til1200~d and \til600~d.  We folded both datasets on these periods.
The folded light curves indicated that the \til600~d peak is a
harmonic of the \til1200~d peak.  We note that the value of 1200~d is
close to half the length of the data.  The light curves of \exo\ are
dominated by the maxima at MJD~2449500 and MJD~2450750.  Therefore,
the 1200~d peak is most likely due to the  non-periodic variations
characteristic of Be stars, rather than a true periodicity.

We find no evidence for any short-term (1--100~d) periodicites in the
light curves of \exo.  The peaks at 1~d in the LS and PDM periodograms
are probably due to the sampling of the datasets.  We find an upper
limit of 0.041~mag semi-amplitude for the presence of any short-term
periodicity in the {\it V}-band data, and 0.047~mag for the {\it
R}-band data.

\section{H\thinspace0544--665}
\label{sect:ho_ch7}

\ho\ was discovered with the {\it HEAO-1} scanning modulation
collimator by  Johnston, Bradt \& Doxsey (1979).  The brightest star
in the error region was found to have a magnitude of B\til16 (star 1
in Figure 6 of Johnston et al.\ 1979), and was later found to be
variable (Thorstensen \& Charles 1980).  van der Klis et al.\ (1983)
found that the star was variable on a timescale of weeks and
determined a spectral type of B0--1 V.  By plotting {\it V} as a
function of {\it B--V}, van der Klis et al.\ (1983) found a
colour-magnitude correlation in which the star becomes redder as it
brightens.  This is typical of Be stars whose variability is due to
variations in the circumstellar disc.  Coe et al.\ (1997) suggested
that, due to the lack of IR emission from the candidate proposed by
van der Klis et al.\ (1983), another star was the optical counterpart
(Star 22 in Coe et al.\ 1997).  However, the Be nature of star 1 of
Johnston et al.\ (1979) was subsequently confirmed by Stevens et al.\
(1999) from spectra of the source in which double peaked H$\alpha$ and
weak H$\beta$ emission is evident.  They therefore concluded that this
star is the optical counterpart of \ho.

\subsection{Light curve}
\label{sect:ho_lc}

Fig.~\ref{fig:ho_lc} shows the MACHO project observations of \ho\
taken  during the period 1993 January 19 to 1997 January 30.  The
magnitudes were  calculated using the absolute calibrations of the
MACHO fields (see  \S~\ref{sect:macho} for details regarding the
acquisition and reduction).

\begin{figure*}
\begin{center}
%\leavevmode
\resizebox*{.7\textwidth}{.32\textheight}{\rotatebox{-90}{\includegraphics{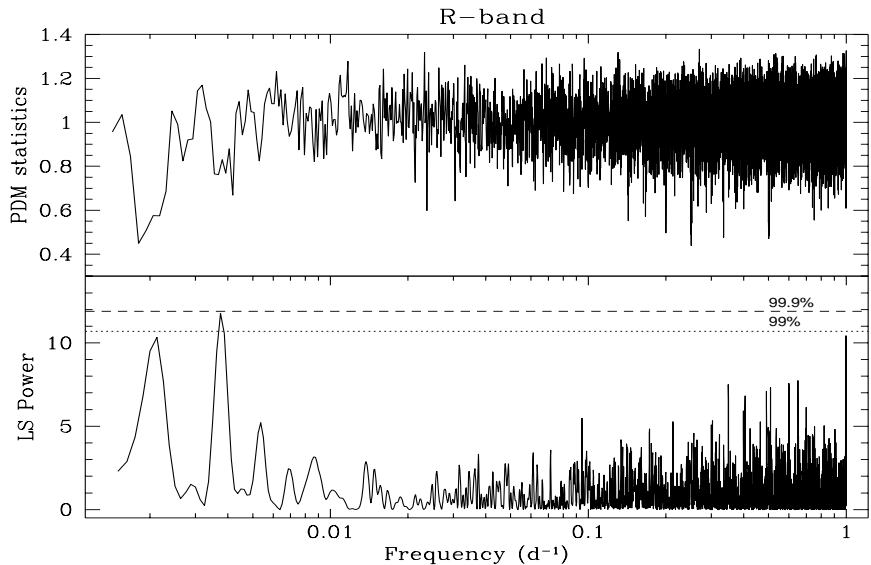}}}
\caption{PDM (top) and LS periodograms (bottom) for the  {\it R}-band
data of \ho, frequency range and resolution is 1.43x$10^{-3}$--1.0
cycle d$^{-1}$ and 1.25x$10^{-4}$ cycle d$^{-1}$, respectively.  The
dotted line is the 99\% confidence level, the dashed line is the
99.9\% confidence level.}\label{fig:ho_pdm}
\end{center} 
\end{figure*}

\subsection{Period analysis}
\label{sect:ho_search}

We detrended the {\it V}- and {\it R}-band data of \ho\ by subtracting
a linear fit.  We then searched for periodicities in the detrended
light curves.  The temporal analysis was performed over  a frequency
range of 1.43x$10^{-3}$--1.0 cycle d$^{-1}$, with a resolution of
1.25x$10^{-4}$ cycle d$^{-1}$.

The LS and PDM periodograms for the {\it R}-band data are shown in
Fig.~\ref{fig:ho_pdm}.  The results for the {\it V}-band data were less
significant.  There is one LS peak which lies above the 99\%
confidence level, however, it is not significant at the 99.9\% level.
We conclude that there are no significant periodicities present in the
light curves of \ho.  We determine an upper limit for detection of any
short-term (1--100~d) periodicity in the {\it V}-band light curve of
0.054~mag semi-amplitude, and 0.075~mag semi-amplitude for the {\it
R}-band light curve of \ho.

\section{Colour variations}
\label{sect:col}

\begin{figure*}
\begin{center}
%\leavevmode
\resizebox*{.55\textwidth}{.23\textheight}{\rotatebox{-90}{\includegraphics{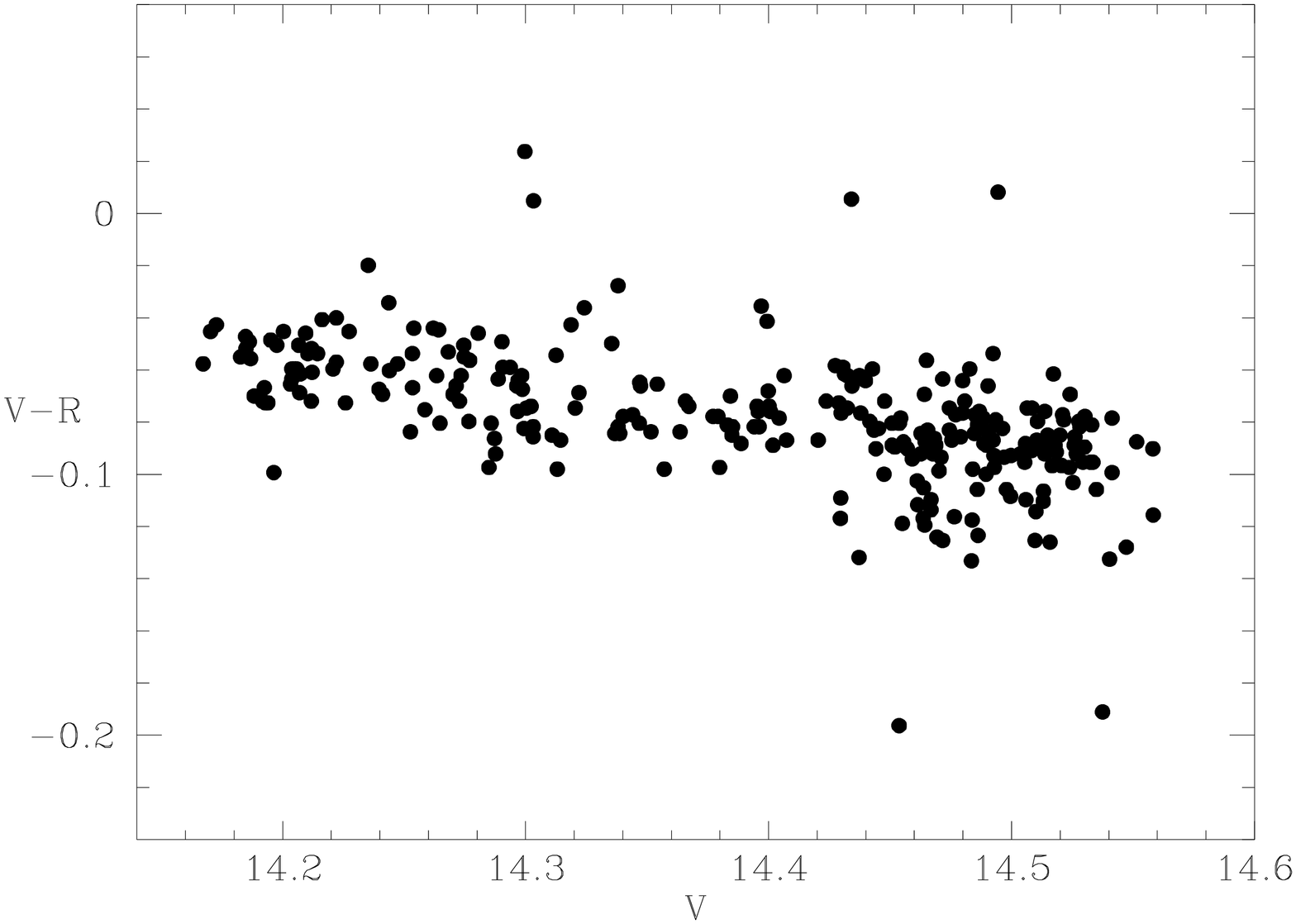}}}
\resizebox*{.55\textwidth}{.23\textheight}{\rotatebox{-90}{\includegraphics{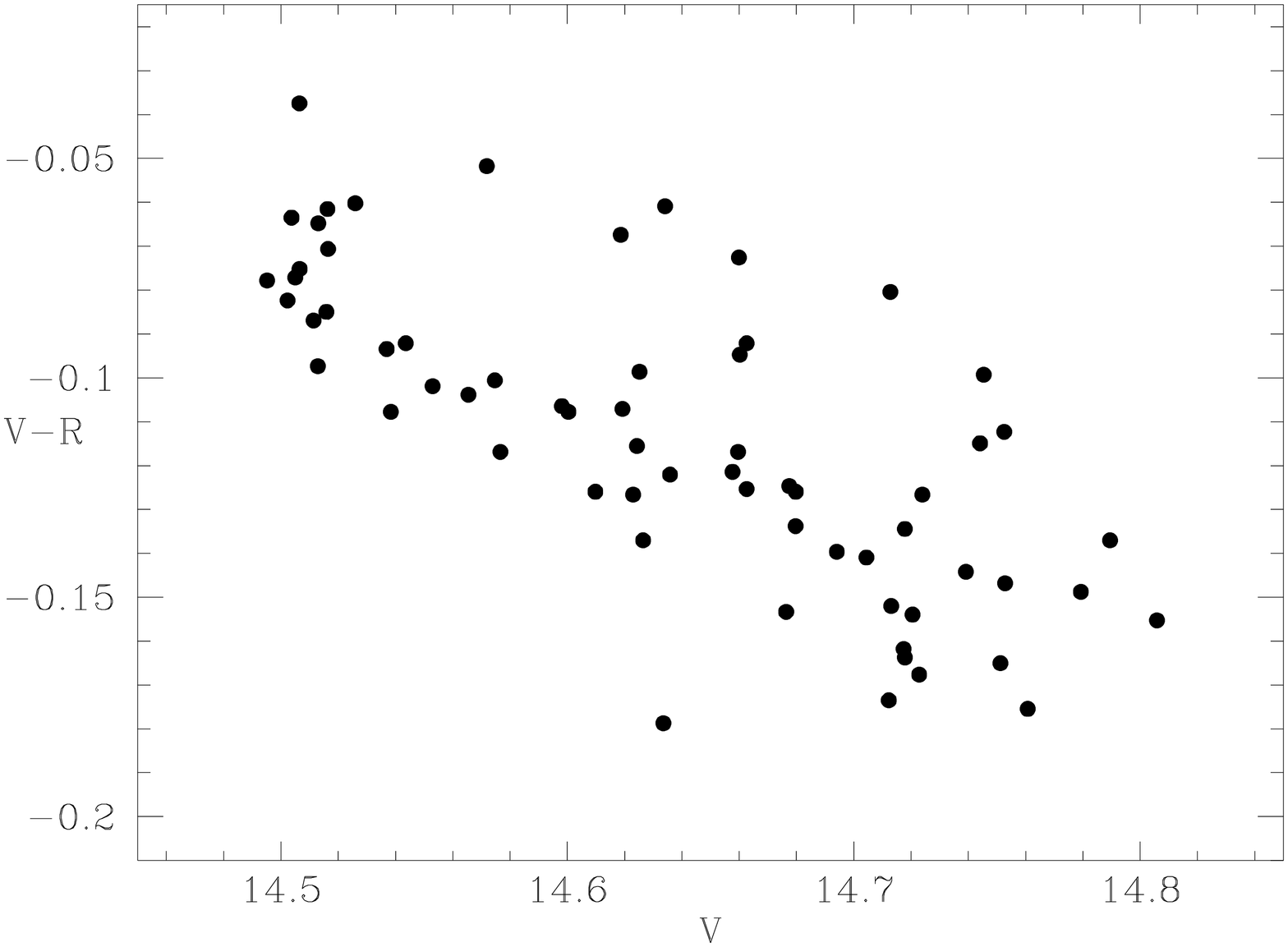}}}
\caption{{\it V} vs {\it V--R} plot for \exo\ (top panel) and \ho\
(bottom  panel).  Note that both sources get redder as they get
brighter.}\label{fig:col}
\end{center} 
\end{figure*}

We constructed {\it V}/({\it V--R}) diagrams for \exo\ and \ho\ to
investigate the changes in colour as the brightness of the sources
vary.  From Fig.~\ref{fig:col} we can see that as \exo\ and \ho\
brighten their colour becomes redder.  This confirms the
colour-magnitude correlation found for \ho\ by van der Klis et al.\
(1983).  This indicates that the variations in the light curves are
likely to be due to variations in the structure of the equatorial disc
around the Be star.  This is due to the disc being redder in {\it
B--V} (i.e.\ cooler) than the Be star (Janot-Pacheco, Motch \& Mouchet
1987).  Thus, the formation of the equatorial disc will increase the
optical brightness of the system by the addition of red light, or it
will make the system appear fainter by masking the Be star, behaviour
which is dependent on the inclination of the system.

\section{Spectroscopy}
\label{sect:spec}

Spectra of the northern component of the double proposed as the
optical counterpart of \exo\ were obtained on the nights of 1998
December 9 and 1998 December 14, using the 1.9~m telescope at SAAO.
The detector was a SITe CCD attached to the Grating Spectrograph,
using a 1200~line/mm grating, giving a resolution of 1 \ang.  Two
900~s exposure spectra of the source were obtained on both nights,
together with Cu-Ar arc spectra for wavelength calibration and dome
flats.  Spectra of \ho\ (Star 1 of Johnston et al.\ 1979) were
obtained on the nights of 1998 December 11 and 1998 December 12,
exactly as for \exo.

The spectra were reduced using {\scshape iraf} in a completely
standard way. The one-dimensional spectra were extracted from the
two-dimensional image using a FWHM aperture to ensure optimum
signal-to-noise ratio. As no flux standards were observed, the spectra
are presented in raw counts only.

\subsection{EXO\thinspace0531--66}

\begin{figure*}
\begin{center}
%\leavevmode
\resizebox*{.65\textwidth}{.3\textheight}{\rotatebox{-90}{\includegraphics{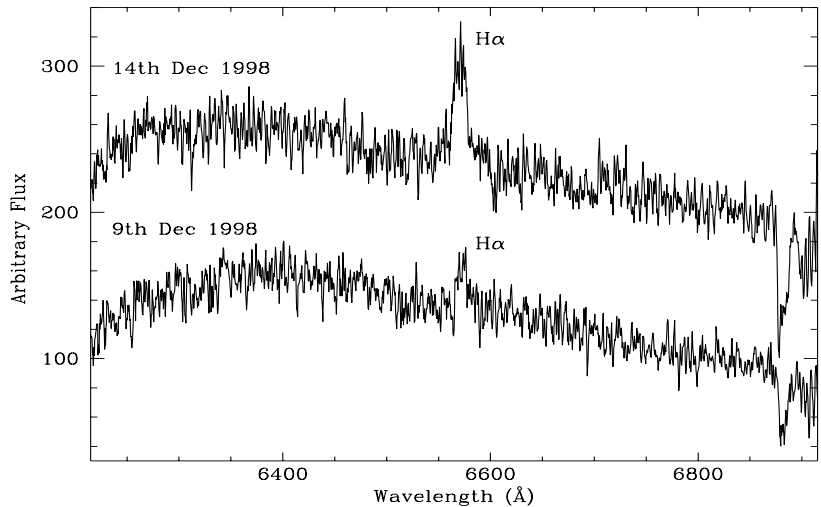}}}
\caption{Combined spectra of the Northern component of the double
proposed as the optical counterpart of \exo\ for the nights of 1998
December 9 (lower) and 1998 December 14 (upper).  The spectra have
been smoothed and offset for clarity.   The H$\alpha$ emission line is
marked, note the increase in emission from the  first night to the
second.}\label{fig:exo_spec}
\end{center} 
\end{figure*}

The summed spectra of \exo\ for each night are shown in
Fig.~\ref{fig:exo_spec} and confirm the presence of H$\alpha$ emission
as seen by Stevens et al.\ (1999).  It is clear that the H$\alpha$
emission profile is changing with time, as shown by the factor \til2.3
increase in flux from the first night to the second.  Balmer emission
in the spectra of Be/XRBs is believed to originate in the
circumstellar material disc around the Be star (Stevens et al.\
1999).  The variations in H$\alpha$ we observe are most likely due to
variations in this disc.

\subsection{H\thinspace0544--665}

\begin{figure*}
\begin{center}
%\leavevmode
\resizebox*{.65\textwidth}{.3\textheight}{\rotatebox{-90}{\includegraphics{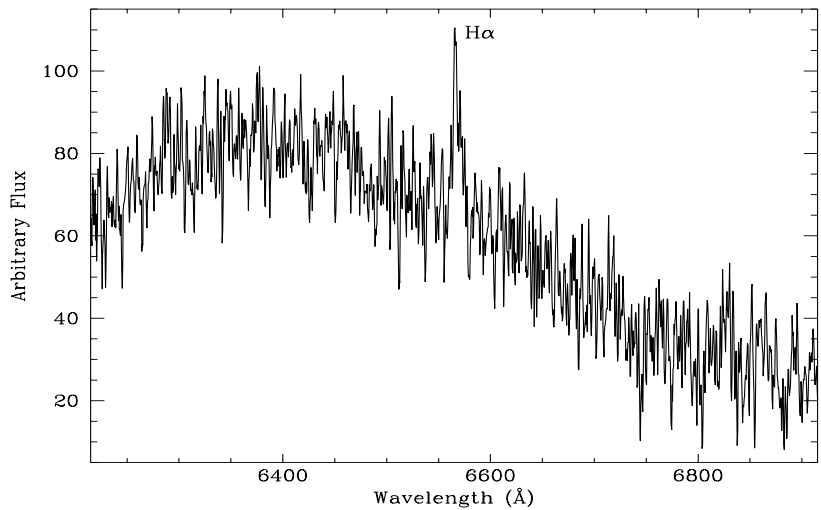}}}
\caption{Combined, smoothed spectrum of \ho.  The H$\alpha$ emission
line is  marked.}\label{fig:ho_spec}
\end{center} 
\end{figure*}

Due to the poorer quality of the spectra of \ho\ we summed all four
spectra in order to increase the signal-to-noise
(Fig.~\ref{fig:ho_spec}).  H$\alpha$ emission is evident in the
combined spectrum, again confirming the results of Stevens et al.\
(1999).

\section{Discussion}
\label{sect:disc}

To investigate the origins of the brightness variations that we
observe in \exo\ and \ho\ it is instructive to consider another
Be/XRT, A0535+26.  This source has been extensively studied since its
discovery in 1975 (Coe et al.\ 1975; Rosenberg et al.\ 1975).  The
system contains a neutron star with a 104~s spin-period in an
elliptical orbit of period \til110.3~d around its O9.7~IIIe companion
(Priedhorsky \& Terrell 1983; Nagase 1989).

Clark et al.\ (1999; hereafter C99) analysed 15 years of optical
photometry of  A0535+26 and compared the variability found with that
of the X-ray data taken  with {\it BATSE} over the same period.  As
the neutron star accretes material from the circumstellar disc around
the Be star, a study of the long-term variability of the envelope can
help in the understanding of the source's X-ray behaviour.  A0535+26
was found to be highly variable in the optical over the 15 year
period.

C99 performed period searches of the long-term optical light curve of
A0535+26.  The analysis failed to find any evidence for modulation of
the circumstellar envelope at the orbital period identified from the
X-ray light curve.  However, periods of \til1400, \til470 and
\til103~d were found in the optical light curve, but their cause and
whether they are coherent over time is unclear.  C99 suggest that the
changes in the optical light curve are due to the variability in the
emission measure of an optically thin circumstellar envelope emitting
via free-free and bound-free emission mechanisms.   As there was no
evidence for optical modulation at the orbital period, C99 concluded
that it is unlikely that the periastron passage of the neutron star
effects the mass-loss rate from the optical companion.  We note
however that Hutchings (1984) found that the H$\beta$ emission from
A0535+26 was modulated on the 111~d period.

We find little evidence for any periodicities in the light curves of
\exo\ and \ho.  The orbital periods of Be/XRTs are usually determined
from the recurrence period of outbursts, particularly in X-rays.  We
found a stable period of 421~d for A0538--66  (see Alcock et al.\
2001) which we suggest is related to the formation and depletion of
the equatorial disc around the Be star, and is certainly {\it not}
orbital in origin.  However, the long-term modulations we observe in
the optical light curves of \exo\ and \ho\ are most likely due to
non-periodic variations characteristic of Be stars.  If stable
long-term variations are present in the light curves of \exo\ and \ho,
they are occurring on much longer timescales than presented here.

If it is assumed that the orbital plane of the neutron star and the
equatorial plane of the Be star are coincident, then when the neutron
star passes through the dense material close to the Be star at
periastron, an X-ray flare should be observed.  C99 tested this model
using their simultaneous optical and X-ray data of A0535+26.  If the
neutron star is accreting directly from the outflowing circumstellar
disc material, the optical and X-ray light curves should be strongly
correlated.

C99 did not find any positive correlation between the optical and
X-ray data, and concluded that this indicates that the neutron star is not 
accreting directly from the stellar wind.  C99 found  that the  X-ray 
outbursts seemed to occur after a period in which the optical light
curve was seen to fade.  This anti-correlation could represent a discrete
episode of disc-loss which occurs radially, the interaction of the
neutron star with this material then triggers the X-ray emission.
They also observed a case in which an optical outburst was observed with
no accompaning X-ray outburst.  We find a slight brightening in the
optical light curve of \exo\ at times of recorded X-ray outburst 
(see Fig~\ref{fig:exo_lc}; Haberl et al.\ 1995, Burderi et al.\
1998).  The lack of a strong correlation between the optical and X-ray
data could be related to the low optical to X-ray luminosities
observed for the sources.

Janot-Pacheco et al.\ (1987) showed that changes in the emission from
the circumstellar disc lead to a correlation between {\it V} and {\it
(B--V)} for A0535+26.  C99 confirmed that a correlation exists where
the system becomes redder as it brightens, owing to a greater
contribution from the cooler disc.  We find the same correlation for
the MACHO data of \exo\ and \ho.  However, another LMC Be/XRT we have
studied, \ao, becomes redder as it fades (see Alcock et al.\ 2001).
As the {\it V}/{\it (B--V)} correlation is  dependent on the
inclination of the system, this suggests that we are viewing A0535+26,
\exo\ and \ho\ at lower inclinations than \ao\ (see McGowan \& Charles 
2002).

Clark et al.\ (1998) investigated the long-term variability of optical
spectroscopy of A0535+26.  The profile and equivalent width of the
H$\alpha$ emission line were found to vary considerably over the 7
year period  of study.  It varied between a single-peaked and
double-peaked or asymmetric structure, and the relative intensities of
the symmetric and asymmetric profiles varied over time.  The
asymmetries were characterised by an additional feature in the blue
or red shoulder, which was sometimes displaced far enough from the
rest wavelength to appear as a separate peak.  Clark et al.\ (1998)
conclude that the variability in the H$\alpha$ line reflects the
changes occurring in the circumstellar envelope, and suggest that the
asymmetries observed in the profile are due to the envelope being
asymmetric in geometry or density.

\section{Conclusions} 
\label{sect:conc}

We can confirm the optical identification of \exo\ with the northern
component  of the close double from the variability in the light
curve, and the H$\alpha$ emission that is observed in the spectra.
The variability in the emission may be due to changes occurring in
the circumstellar envelope as for A0535+26 (Clark et al.\ 1998).  Our
\ho\ spectrum also shows H$\alpha$ emission, but more spectra are
required in order to investigate its variability.

Haberl et al.\ (1995) proposed an orbital period for \exo\ of
600--700~d.  Dennerl et al.\ (1996) proposed a much shorter period for
the source in the range of 4--70~d.  We find no variability on these 
timescales in the MACHO data.  We find that the upper limits for a
short-term (1--100~d) periodic modulation to be detected in the 
{\it V}- and {\it R}-band light curves of \exo\ are 0.041~mag and 0.047~mag,
respectively, and 0.054~mag and 0.075~mag, respectively, for \ho.

The colour changes for A0535+26, \exo\ and \ho\ are consistent with
the systems becoming redder as they brighten owing to the greater
contribution from the cooler equatorial disc (Janot-Pacheco et al.\
1987).  This provides evidence that we are observing the systems at a
lower inclination than A0538--66 (see McGowan \& Charles 2002).  We
detected the orbital period in the optical light curve of A0538--66 
(see Alcock et al.\ 2001) which is also seen in X-rays.  C99 find no
evidence for modulation of the optical light curve of A0535+26 on the
X-ray-determined orbital period.  Hence, the system inclination may
determine whether the orbital period of a Be system is detectable
optically, and therefore we may only be able to detect those of \exo\
and \ho\ in X-rays.  However, the detection of the
orbital period in both optical and X-rays for A0538--66 may be
connected to its short orbital period (16.6~d).  If
this is true, it suggests that the orbital periods for
\exo\ and \ho\ are both longer than \til20~d.

\section{Acknowledgements}

We thank Thebe Medupe for generously obtaining the optical
spectra with the SAAO 1.9~m telescope.  
This paper utilizes public domain data obtained by the MACHO project,
jointly funded by the US Department of Energy through the University
of California, Lawrence Livermore National Laboratory under contract
No. 7405-Eng-48, by the National Science Foundation through the Center
for Particle Astrophysics of the University of California under
cooperative agreement AST-8809616, and by the Mount Stromlo and Siding
Spring Observatory, part of the Australian National University.

\label{lastpage}

\end{document}